\documentclass{JFM-FLM_Au}

\usepackage{amsmath}
\usepackage{bm}
\usepackage{physics}

\lefttitle{B.L. Espa\~nol, J.C. Vassilicos, R. Zamansky, M. Obligado and P.D. Mininni}
\righttitle{Journal of Fluid Mechanics}

\title{Emergence of non-equilibrium turbulent dissipation laws from Lagrangian trajectories}

\author{B.L. Espa\~nol\aff{1,2,3}, J.C. Vassilicos\aff{4}, R. Zamansky\aff{5}, M. Obligado\aff{4,6} \and P.D. Mininni\aff{1,2,3}}

\affiliation{\aff{1}Universidad de Buenos Aires (UBA), Facultad de Ciencias Exactas y Naturales, Departamento de F\'isica, Ciudad Universitaria, 1428 Buenos Aires, Argentina
\aff{2}CONICET-UBA, Instituto de F\'isica Interdisciplinaria y Aplicada (INFINA), Ciudad Universitaria, 1428 Buenos Aires, Argentina
\aff{3}CNRS-CONICET-UBA, Institut Franco-Argentin de Dynamique des Fluides pour l'Environnement (IFADyFE), IRL2027, Ciudad Universitaria, 1428 Buenos Aires, Argentina
\aff{4}Univ.~Lille, CNRS, ONERA, Arts et Metiers Institute of Technology, Centrale Lille, UMR 9014-LMFL-Laboratoire de M\'ecanique des Fluides de Lille - Kamp\'e de F\'eriet, F-59000 Lille, France
\aff{5}Institut de M\'ecanique des Fluides de Toulouse (IMFT), CNRS-INPT-UPS, Toulouse, France
\aff{6}Institut Universitaire de France (IUF), Paris, France}

\corresau{B.L. Espa\~nol, \email{bespanol@df.uba.ar}}

\begin{document}
\maketitle

\begin{abstract}
In unsteady turbulent flows, the dimensionless dissipation rate and the Taylor-scale Reynolds number obey a non-equilibrium law, a signature of the finite response time of the energy cascade. This law describes the flow globally, and how it emerges from the elementary degrees of freedom of turbulence remains an open question. We address this question using Lagrangian tracers in numerical simulations of homogeneous and isotropic turbulence, whose global state undergoes spontaneous, slow excursions around its mean that probe the non-equilibrium law with no externally imposed unsteadiness. We show that local analogues of the dissipation rate and of the Reynolds number, evaluated pointwise either along Lagrangian trajectories or at fixed points in the Eulerian field, follow a power law of purely statistical nature. The power law is fixed by the one-point statistics of velocity and dissipation alone, and its slope measures the intensity of dissipation fluctuations. By averaging over an increasing number of trajectories, we then construct a family of observables that interpolates between local and global behavior, and show that the global law emerges from an average over the fluctuations. The results show that the relation between dissipation and Reynolds number depends on the scale, with the fluctuations of the intermittent turbulent field dominating locally, and the cascade-enforced coupling between energy and dissipation dominating globally.
\end{abstract}

\section{Introduction}

The dimensionless dissipation rate, $C_\varepsilon = \varepsilon L / u_{\rm rms}^3$ (where $\varepsilon$ is the energy dissipation rate, $u_{\rm rms}$ is the root-mean-square (r.m.s.) velocity, and $L$ is the integral scale), plays a central role in turbulence theory. The relation $\varepsilon \sim u_{\rm rms}^3/L$ (i.e., $C_\varepsilon$ is a constant of order unity) has long been regarded as one of the cornerstone assumptions of the field \citep{Vassilicos_2015}: it states that the rate at which turbulence dissipates energy is set by the large scales alone, independently of the viscosity, and in this form it enters most turbulence theories and models. And at sufficiently large Reynolds numbers, for flows consistent with Kolmogorov phenomenology, the long-time average of $C_\varepsilon$ indeed seems to approach a constant, the hallmark of the Richardson-Kolmogorov equilibrium cascade, in which the mean inertial-range inter-scale energy transfer balances the mean energy dissipation rate $\varepsilon$ \citep{Sreenivasan_1997, Vassilicos_2015}.

This equilibrium picture, however, is a statement about long-time averages of statistically stationary homogeneous turbulence (in the sense of being taken over several eddy
turnover times of the flow). Instantaneously, the large scales of a turbulent flow evolve: trivially so in decaying flows, but also in flows that are statistically steady over long times, whose global quantities display fluctuations around their mean values \citep{Goto_2015, Zapata_2024}. As the large scales evolve, the energy they inject must still traverse the cascade before being dissipated, and the dissipation responds with a delay of the order of the eddy turnover time \citep{Goto_2015, Goto_2016, Zapata_2024}. The resulting scale-by-scale energy imbalance breaks the constancy of $C_\varepsilon$: the dimensionless dissipation instead follows the non-equilibrium law $C_\varepsilon \sim \sqrt{\Rey_0}/\Rey_\lambda$, where $\Rey_\lambda$ is the Taylor-scale Reynolds number and $\Rey_0$ a flow-dependent global (and constant) Reynolds number. This law has been observed across a wide range of unsteady flows, decaying as well as forced \citep{Valente_2014, Goto_2015, Vassilicos_2015, Goto_2016, Vassilicos_2023, Ghira_2026, Ahmad_2026}, including statistically steady flows whose spontaneous or externally forced large-scale excursions have been shown to drive extreme clustering of inertial particles \citep{Zapata_2024}.

However, this non-equilibrium law is still a statement about the flow as a whole, built from averaged quantities---i.e., integrated spatially over volumes comparable to the integral length scale. Recently, several studies have asked whether it survives at smaller scales within the inertial range, evaluating the dissipation rate and the Reynolds number locally in an Eulerian frame---over short time windows along single probes in grid-turbulence experiments \citep{Zheng_2023}, or over spatial subdomains still of the order of the integral scale in simulations of incompressible \citep{Nishimoto_2026a} and compressible \citep{Nishimoto_2026b} turbulence---and have reported that the same $\Rey_\lambda^{-1}$ scaling holds, suggesting the possibility of non-equilibrium, local, relations. At the smallest scales within the inertial range, however, a different physics takes over: the dissipation is strongly intermittent, with approximately lognormal fluctuations spanning orders of magnitude \citep{Kolmogorov_1962, Obukhov_1962, Sreenivasan_1997, Donzis_2008}, and pointwise quantities bear little resemblance to their spatial means. How the whole-flow dissipation law emerges from the elementary degrees of freedom of the turbulence, and at what scale it does, remains an open question.

The Lagrangian frame provides a natural way to address this question: fluid elements sample turbulence along their trajectories, and can be aggregated at will into probes of the turbulence at any scale \citep{Pumir_2000, Toschi2009, Espanol_2026}. Here we use Lagrangian tracers in direct numerical simulations of homogeneous and isotropic turbulence to follow the relation between dissipation and Reynolds number across scales, from the individual fluid element to the flow as a whole. The flow is statistically steady over long times, but its global state undergoes spontaneous, slow excursions around its mean, and it is through these excursions, rather than through any externally imposed protocol, that the non-equilibrium law manifests itself. We show that at the level of individual fluid elements, the relation between $C_\varepsilon$ and $\Rey_\lambda$ is given by a steep power law of purely statistical origin, whose exponent is fixed by the one-point statistics of velocity and dissipation---in particular, by the intensity of the fluctuations of the latter---and contains no cascade dynamics. Being a single-point property, the local law found along trajectories also holds for point measurements in the Eulerian field. Averaging over groups of tracers of increasing number then reveals how the dynamical law emerges: the measured exponent abandons this statistical behaviour at a crossover group size that can be obtained from a balance between sampling and temporal fluctuations. Which physics dominates in the $(\Rey_\lambda, C_\varepsilon)$ plane---the fluctuations of the intermittent turbulent field or the delayed dynamics of the cascade---thus depends on the scale considered.

\section{Methods}

We performed direct numerical simulations (DNSs) of the incompressible Navier--Stokes equation, using an external forcing $\boldsymbol{F}$ to sustain a turbulent regime. The equations are solved in a three-dimensional cubic $(2\upi L_0)^3$-periodic domain, where $L_0$ is a unit length, using the parallel pseudospectral code GHOST \citep{Mininni_2011c, Rosenberg_2020}, with a spatial resolution of $512^3$ grid points. The flow is forced at the largest scales by
\begin{equation}
	\boldsymbol{F} = F_0 \sum_{\abs{\boldsymbol{k}} \in (0, 2]} \Real
	\left\{ \frac{\mathrm{i}\, \boldsymbol{k} \times \hat{\boldsymbol{e}}}{\abs{\boldsymbol{k}}}\, \mathrm{e}^{\mathrm{i}\,(\boldsymbol{k} \bcdot \boldsymbol{r} + \phi_{\boldsymbol{k}})} \right\},
\end{equation}
where $F_0$ is a constant amplitude, $\boldsymbol{k}$ the wave vector, $\hat{\boldsymbol{e}}$ a unit vector, $\phi_{\boldsymbol{k}}$ a random phase, and $\Real\{\cdot\}$ denotes the real part. The forcing injects energy only at the largest scales, which set the integral scale of the flow, and the random phases of each mode evolve continuously in time, with a correlation time of $0.5$ large-scale eddy turnover times. The forcing amplitude and the kinematic viscosity were adjusted so that the r.m.s.~velocity is $U_0$ and of order unity, and the simulation is well resolved with $\kappa\eta > 1$, where
$\kappa = 512/3$ is the maximum resolved wavenumber and $\eta$ the Kolmogorov length scale. The global Taylor-scale Reynolds number is defined as
\begin{equation}
	\Rey_\lambda = \sqrt{\frac{15}{\nu \ev{\varepsilon}}} \ev{u^2},
	\label{eq:relam_def}
\end{equation}
where $\ev{\cdot}$ denotes a spatial average over the entire flow, and $u^2 = (u_x^2 + u_y^2 + u_z^2)/3$ is the squared velocity averaged over the three components; in the simulations $\Rey_\lambda \approx 215$. The integral scale of the flow is
obtained from the instantaneous energy spectrum $E(k, t)$ as
\begin{equation}
	L(t) = 2\upi\, \frac{\int E(k, t)\, k^{-1}\, \dd k}{\int E(k, t)\, \dd k},
	\label{eq:integral_scale}
\end{equation}
and its time average $L = \ev{L(t)}_t$ allows us to define the large-scale eddy turnover time $T_0 = L/U$, with
$U = \ev{u^2}^{1/2}$.

Once the velocity field reaches the statistically steady turbulent regime, we seed the flow with $N = 5\times10^5$ Lagrangian tracers at random positions, and integrate their trajectories for approximately $27\, T_0$ following the equation $\dv*{\boldsymbol{x}_p}{t} = \boldsymbol{u}(\boldsymbol{x}_p, t)$, where $\boldsymbol{x}_p$ is the particle position at time $t$, and $\boldsymbol{u}(\boldsymbol{x}_p, t)$ the fluid velocity at that position. The tracers possess no inertia and follow the flow within numerical errors, acting as a set of sensors that sample the turbulent field along fluid element trajectories. All local quantities are evaluated along the trajectories by interpolating the corresponding Eulerian fields to the instantaneous particle positions using high-order spline methods \citep{Yeung_1988}.

\section{The global non-equilibrium law}
\label{sec:global}

\begin{figure}
	\centering
	\includegraphics[width=0.95\linewidth]{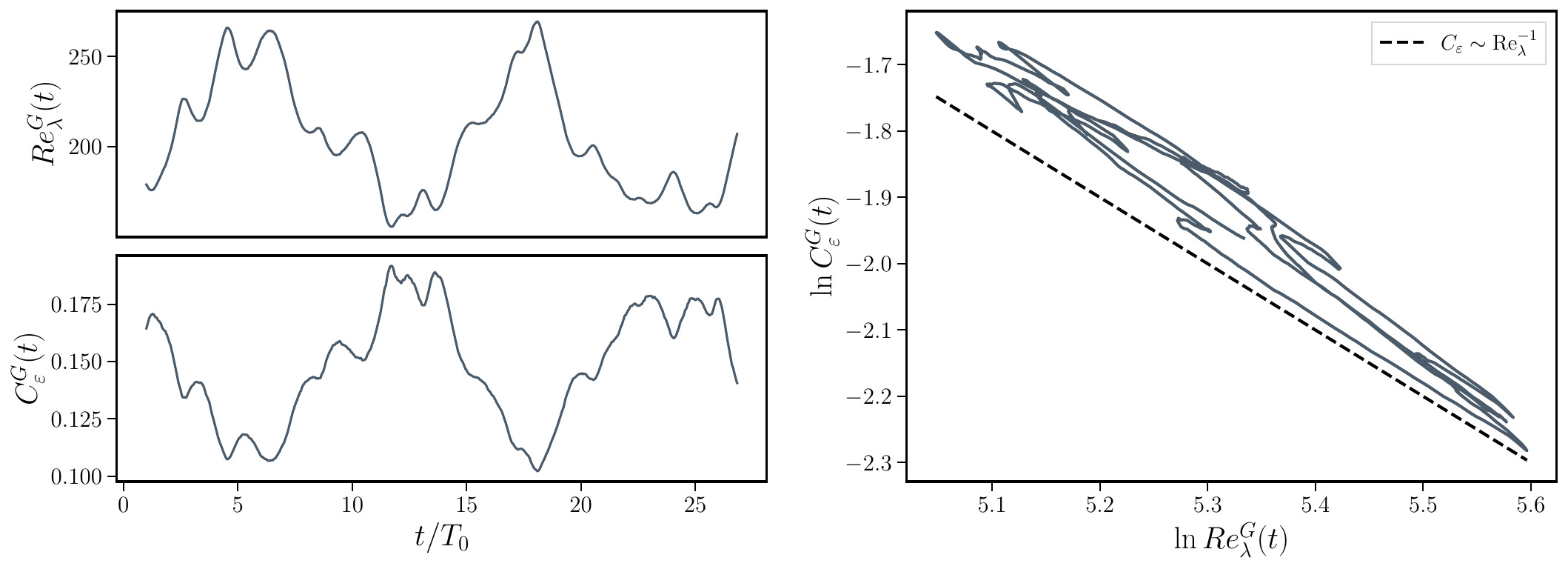}
	\caption{Global Taylor-based Reynolds number $\Rey_\lambda^G$ and global dimensionless dissipation rate $C_\varepsilon^G$, computed from equation~\eqref{eq:global_defs} using all tracers as sensors. The left panels show $\Rey_\lambda^G$ (top) and $C_\varepsilon^G$ (bottom) as a function of time, in units of the turnover time $T_0$. The right panel shows the same evolution in the $(\Rey_\lambda^G, C_\varepsilon^G)$ plane, where each point along the curve corresponds to a different time in the left panels. The dashed line indicates the power law in the global non-equilibrium dissipation law, $C_\varepsilon \sim \Rey_\lambda^{-1}$, as a reference.}
	\label{fig:global}
\end{figure}

We start by characterizing the global unsteadiness of the flow. To this end we use the tracers as a set of sensors distributed over the domain: at each time, we average the squared fluid velocity and the local energy dissipation rate over all particle positions, and compute the global quantities
\begin{equation}
	\Rey_\lambda^G(t) = \sqrt{\frac{15}{\nu \ev{\varepsilon}_N}} \ev{u^2}_N \qc
	C_\varepsilon^G(t) = \frac{\ev{\varepsilon}_N L(t)}{\ev{u^2}_N^{3/2}},
	\label{eq:global_defs}
\end{equation}
where $\ev{\cdot}_N$ denotes the average over the positions $\boldsymbol{x}_p$ of all $N$ tracers at a fixed time, and $u^2(\boldsymbol{x}_p, t)$ and $\varepsilon(\boldsymbol{x}_p, t)$ are respectively the squared fluid velocity and the local energy dissipation rate evaluated at those positions. We verified that the number of tracers used ($N = 5 \times 10^5$) is enough to converge all unsteady statistics reported in the present work. As tracers sample the flow homogeneously, $\langle u^2 \rangle_N^{1/2}$ converges to the r.m.s.~turbulent velocity and $\langle \varepsilon \rangle_N$ to the mean dissipation, and equation~\eqref{eq:global_defs} reduces to the usual definitions of the Taylor-based Reynolds number and of the dimensionless dissipation rate. Figure~\ref{fig:global} (left) shows $\Rey_\lambda^G$ and $C_\varepsilon^G$ as a function of time, in units of the turnover time $T_0$. Both quantities display excursions with a characteristic time scale of the order of the large-scale eddy turnover time, and are clearly anticorrelated: when turbulence becomes stronger (i.e., when $\Rey_\lambda^G$ grows), the global normalized dissipation $C_\varepsilon^G$ decreases, as previously reported for this kind of flows \citep{Vassilicos_2015, Goto_2015, Zapata_2024, Ahmad_2026}.

The anticorrelation can be quantified by following the system in the $(\Rey_\lambda^G, C_\varepsilon^G)$ plane, as shown in figure~\ref{fig:global} (right), where each point along the curve corresponds to a different time. Over the range of $\Rey_\lambda^G$ accessible to a single flow, with excursions of about $\pm 25\%$ around the mean, the data are compatible with the power law $C_\varepsilon^G \sim (\Rey_\lambda^G)^{-1}$. This behaviour is in agreement with the non-equilibrium Eulerian dissipation law $C_\varepsilon \sim \sqrt{\Rey_0}/\Rey_\lambda$ reported in a variety of unsteady turbulent flows \citep{Valente_2014, Goto_2015, Vassilicos_2015, Vassilicos_2023, Ahmad_2026} and compatible with results in \citet{Zapata_2024} where simulations with different viscosities (and $\Rey_0$) were compared. Here, $\Rey_0$ is a global Reynolds number, constant for a given flow: it sets the prefactor of the law, and the excursions of a single flow can therefore only probe the $\Rey_\lambda^{-1}$ dependence. Over our time series the best fit for the slope is $-1.12$, slightly steeper than $-1$. There are two possible reasons for this small discrepancy: lack of statistics, as we only consider 27 turnover times, or the fact that the present flow, although unsteady at turnover timescales, is statistically steady over longer timescales: its excursions do not evolve freely as in decaying turbulence, and are bounded to eventually relax back to the mean state, a constraint that the non-equilibrium law originally did not incorporate. Note that the non-equilibrium law differs from the classical prediction of the Richardson--Kolmogorov equilibrium cascade, $C_\varepsilon = \text{const}$: in unsteady flows there is a scale-by-scale energy imbalance, and the energy dissipated at the small scales does not adjust instantaneously to the energy injected at the large scales \citep{Goto_2015, Goto_2016}. The same imbalance is visible in the loops described by the system in figure~\ref{fig:global} (right): the curve is not a single-valued function of $\Rey_\lambda^G$, as the dissipation responds to changes in the large scales with a delay of the order of the turnover time \citep{Goto_2016, Zapata_2024}. The law also has a simple geometrical interpretation. From equation~\eqref{eq:global_defs}, $C_\varepsilon^G \Rey_\lambda^G = 15 L/\lambda^G$, where $\lambda^G = (15 \nu \langle u^2 \rangle_N / \langle \varepsilon \rangle_N)^{1/2}$ is the global Taylor microscale; the $-1$ scaling is thus equivalent to the statement that the scale separation $L/\lambda^G$ remains frozen while the flow goes through its excursions \citep{Vassilicos_2015, Goto_2016}. In what follows it will prove useful to read this law as a statement about the joint fluctuations of velocity and dissipation: $\langle u^2 \rangle_N(t)$ and $\langle \varepsilon \rangle_N(t)$ do not fluctuate independently, but co-evolve locked by the cascade, with the dissipation following the energy with a delay of the order of the turnover time; the $-1$ slope, and the loops around it, are the signature of this coupling. Finally, the fact that the tracers recover this law confirms that they correctly capture the global state of the flow.

\section{The Lagrangian non-equilibrium law}

These definitions are intrinsically global: they require averaging over many tracers (or over the entire domain in the Eulerian frame) at a fixed time. However, a fluid element evolves in time, and is not fixed at a given position. Instead, it evolves in time and in space simultaneously, sampling the flow along its trajectory. What Lagrangian relation between dissipation and Reynolds number, if any, is then seen by each fluid element? To answer this question we need local definitions of $\Rey_\lambda$ and $C_\varepsilon$. The natural candidates are
\begin{equation}
	\Rey_\lambda^L(\boldsymbol{x}_p, t) = \sqrt{\frac{15}{\nu \varepsilon}} \, u^2 \qc
	C_\varepsilon^L(\boldsymbol{x}_p, t) = \frac{\varepsilon L(t)}{u^3},
	\label{eq:local_defs}
\end{equation}
where the fluid velocity and the local dissipation are now evaluated at the position of an individual particle, $u = u(\boldsymbol{x}_p, t)$ and $\varepsilon = \varepsilon(\boldsymbol{x}_p, t)$. Note that the integral scale $L(t)$ remains a global quantity, as it characterizes the large-scale structure of the flow, and all tracers are expected to see the same flow correlation length, imposed by the external forcing.

\begin{figure}
	\centering
	\includegraphics[width=0.48\linewidth]{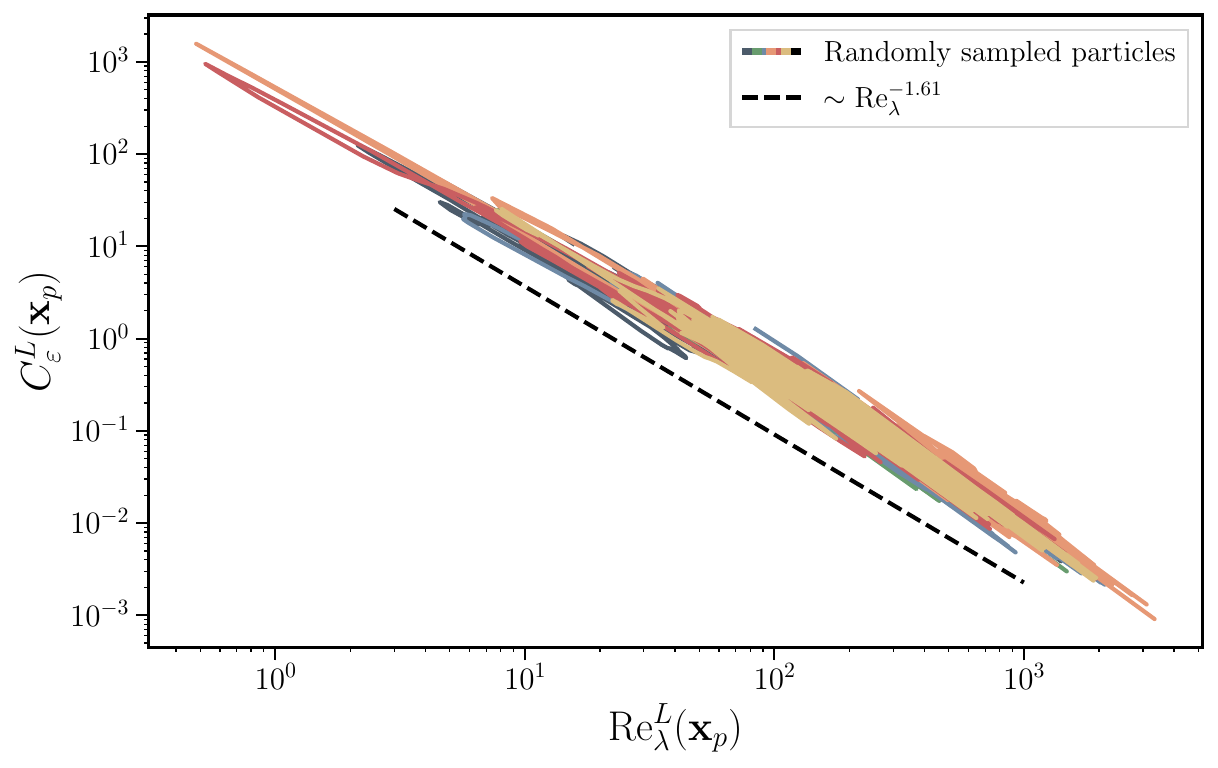}
	\hfill
	\includegraphics[width=0.48\linewidth]{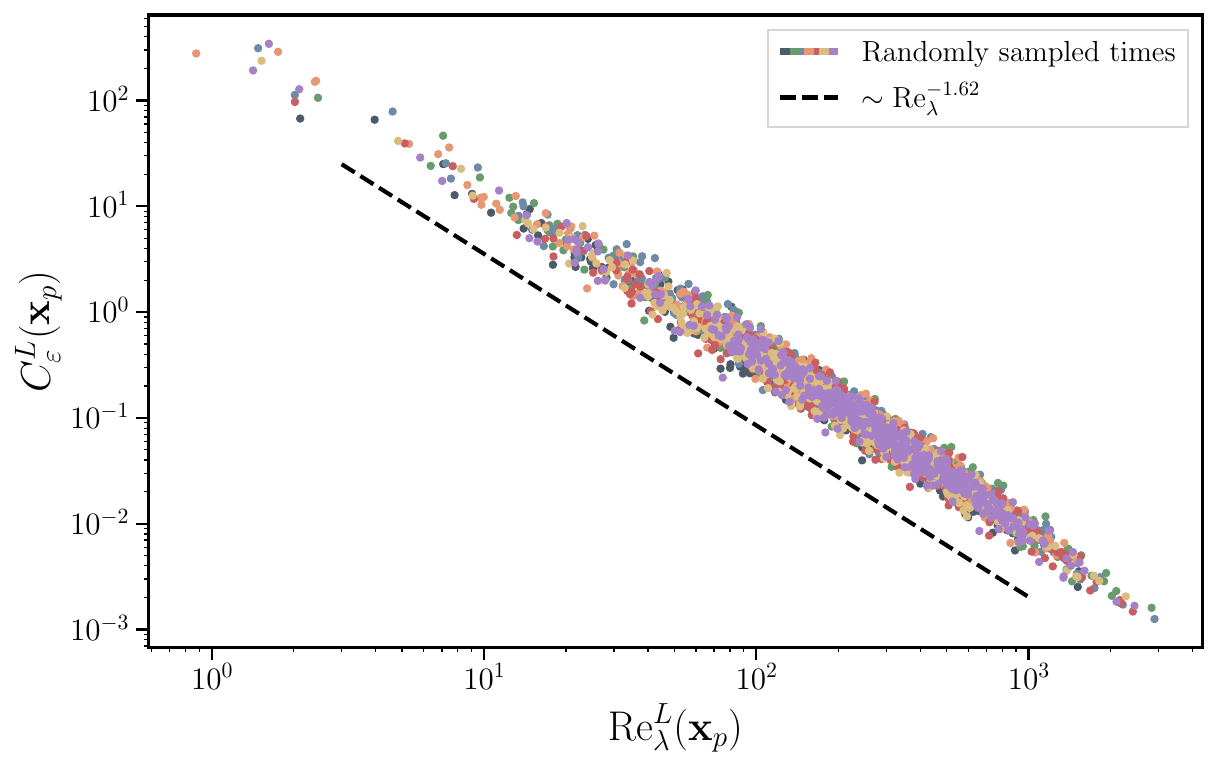}
	\caption{Local (Lagrangian) dimensionless dissipation rate $C_\varepsilon^L$ as a function of the local Taylor-based Reynolds number $\Rey_\lambda^L$, equation~\eqref{eq:local_defs}. The left panel shows a few randomly sampled tracers (one colour per trajectory), where each point along a curve corresponds to a different time along the particle trajectory. The right panel shows a random subset of the tracers at a few randomly sampled fixed times (one colour per time). In each panel, the dashed line indicates the corresponding fitted power law.}
	\label{fig:local}
\end{figure}

Figure~\ref{fig:local} (left) shows $C_\varepsilon^L$ as a function of $\Rey_\lambda^L$ for a few randomly sampled tracers, where each point along a curve corresponds to a different time along the particle trajectory, and each color corresponds to a different tracer. Note that the local (Lagrangian) quantities span several orders of magnitude, in contrast with the modest excursions of their global counterparts in figure~\ref{fig:global} (right). Moreover, the power law followed by the fluid elements is steeper than the global (Eulerian) one: all particles collapse onto the same power law, $C_\varepsilon^L \sim (\Rey_\lambda^L)^{\alpha}$ with $\alpha = -1.61 \pm 0.04$ (mean and standard deviation of the per-trajectory fits over all tracers).

This exponent can be predicted from the one-point statistics of the flow. From equation~\eqref{eq:local_defs} it follows that
\begin{equation}
	C_\varepsilon^L \left(\Rey_\lambda^L\right)^{3/2} = 15^{3/4}\, \frac{L(t)}{\eta^L(\boldsymbol{x}_p, t)},
	\label{eq:identity}
\end{equation}
where $\eta^L = (\nu^3/\varepsilon)^{1/4}$ is a local Kolmogorov scale based on the local dissipation. Equation~\eqref{eq:identity} is an identity that holds point by point, and is the local analogue of the relation $C_\varepsilon \Rey_\lambda \propto L/\lambda$ of equilibrium turbulence \citep{Vassilicos_2015}: in the $(\Rey_\lambda^L, C_\varepsilon^L)$ plane, fluid elements must lie on lines of slope $-3/2$, parametrized by the local scale separation $L/\eta^L$. This slope is close to (albeit not exactly) the one observed. This also explains why the points in figure~\ref{fig:local} collapse onto a narrow band, even though locally $\varepsilon$ spans orders of magnitude: as $L/\eta^L \propto \varepsilon^{1/4}$, fluctuations of the dissipation around this relation enter the scale separation attenuated by the power $1/4$, and the scatter of the points around the line is correspondingly small. Moreover, the definitions constrain how a fluid element can move in this plane. Fluctuations of the velocity at fixed dissipation displace the element along lines of slope exactly $-3/2$. Fluctuations of the dissipation at fixed velocity instead displace it along lines of slope $-2$, because $\varepsilon$ enters both axes at once, with opposite signs and different weights: a fluctuation $\delta \ln \varepsilon$ (with $\delta$ a small real number) shifts $\ln C_\varepsilon^L$ by the full $\delta \ln \varepsilon$ (as $C_\varepsilon^L \propto \varepsilon$) while lowering $\ln \Rey_\lambda^L$ by half of it (as $\Rey_\lambda^L \propto \varepsilon^{-1/2}$), a displacement of slope $1/(-1/2) = -2$. Physically, at fixed velocity a more intense dissipation means steeper velocity gradients, and thus a smaller local Taylor microscale $\lambda^L = u \sqrt{15 \nu / \varepsilon}$: when a fluid element enters an intense dissipative structure, its local Reynolds number $\Rey_\lambda^L = u \lambda^L / \nu$ drops and its normalized dissipation rises in response to the same event. A cloud generated by independent fluctuations of the two types must therefore display a slope between $-3/2$ and $-2$, fixed by their relative amplitude.

The previous argument, however, only approximately bounds $\alpha$ between these two values. We can provide instead a quantitative expression for $\alpha$. Taking the logarithm of equation~\eqref{eq:local_defs}, $\ln C_\varepsilon^L = \ln \varepsilon + \ln L - 3 \ln u$ and $\ln \Rey_\lambda^L = 2 \ln u - (1/2) \ln \varepsilon + \text{const}$, and using that the slope of a linear regression is $\alpha = \text{Cov}(\ln \Rey_\lambda^L, \ln C_\varepsilon^L) / \text{Var}(\ln \Rey_\lambda^L)$, one obtains
\begin{equation}
	\alpha = \frac{-6 \sigma^2_{\ln u} - \sigma^2_{\ln\varepsilon}/2 + (7/2)\, C_{u\varepsilon} + 2 C_{uL} - C_{\varepsilon L}/2}{4 \sigma^2_{\ln u} + \sigma^2_{\ln\varepsilon}/4 - 2 C_{u\varepsilon}},
	\label{eq:slope_general}
\end{equation}
where $\sigma^2_{\ln u}$ and $\sigma^2_{\ln\varepsilon}$ are respectively the variances of $\ln u$ and $\ln \varepsilon$, and $C_{u\varepsilon}$, $C_{uL}$, and $C_{\varepsilon L}$ are the covariances between the pairs of $\ln u$, $\ln \varepsilon$, and $\ln L$. For individual fluid elements variances dominate. Covariances involving $\ln L$ are small because $L(t)$ is the same for all tracers at a given time, and $C_{u\varepsilon}$ is small because, pointwise, velocity and dissipation are only weakly correlated, consistently with \citet{Larssen_2023}. Specifically, in our simulation $\sigma^2_{\ln u} = 0.22$ and $\sigma^2_{\ln\varepsilon} = 1.2$, against $C_{u\varepsilon} = 0.03$ and $C_{uL}$, $C_{\varepsilon L}$ both below $10^{-2}$ in absolute value. Neglecting the covariances, equation~\eqref{eq:slope_general} reduces to
\begin{equation}
	\alpha = -\frac{3}{2} - \frac{\sigma^2_{\ln\varepsilon}/8}{4 \sigma^2_{\ln u} + \sigma^2_{\ln\varepsilon}/4},
	\label{eq:slope}
\end{equation}
which interpolates monotonically between $-3/2$ ($\sigma^2_{\ln\varepsilon} \to 0$) and $-2$ ($\sigma^2_{\ln u} \to 0$): for decoupled fluctuations, no variance budget can take the slope outside the band $[-2, -3/2]$, while a nonzero correlation can. In terms of equation~\eqref{eq:identity}, the correction to $-3/2$ in equation~\eqref{eq:slope} is associated with the fluctuations of $L/\eta^L$. Equation~\eqref{eq:slope} with the variances measured directly from the simulation gives $\alpha = -1.63$, against $-1.61 \pm 0.04$ from the best fit to the data in figure~\ref{fig:local}. The small difference vanishes when the covariances are accounted for in equation~\eqref{eq:slope_general}. At the level of individual fluid elements, we can thus conclude that velocity and dissipation fluctuate in an effectively decoupled way, in sharp contrast with their cascade-locked global co-evolution.

More telling is to estimate $\alpha$ in equation~\eqref{eq:slope} from general properties of turbulent flows, without using values from the simulation. Note that both variances in equation~\eqref{eq:slope} are known quantities in turbulence theory: for a near-Gaussian one-point velocity distribution, the squared velocity magnitude follows a $\chi^2_3$ distribution, so that $\ln u^2$ follows a log-gamma distribution, which fixes $\sigma^2_{\ln u} = \psi_1(3/2)/4 \approx 0.23$ (where $\psi_1$ is the trigamma function), independently of the turbulence intensity. Moreover, the approximately lognormal statistics of the dissipation \citep{Kolmogorov_1962, Obukhov_1962, Pearson_2018} give $\sigma^2_{\ln\varepsilon}$ between $1$ and $1.6$ at the Reynolds number of our simulation \citep{Sreenivasan_1997, Donzis_2008}. Equation~\eqref{eq:slope} then predicts $\alpha$ between $-1.60$ and $-1.65$, well within the uncertainty of the measured exponent. Note that this value can be estimated from theoretical results without any parameter extracted from the simulation.

The departure of the slope from $-3/2$ then becomes a direct, quantitative measure of the width of the logarithmic dissipation fluctuations relative to those of the velocity: in a field with non-fluctuating $\varepsilon$ the points in figure~\ref{fig:local} would align exactly with the constant $L/\eta^L$ lines of slope $-3/2$, and it is the broad fluctuations of $\varepsilon$, several times larger than those of the velocity, that tilt the points in the direction of the dissipation-dominated slope of $-2$. As a consequence, the exponent is not expected to be universal: as the width of the dissipation distribution grows slowly with the Reynolds number \citep{Donzis_2008, Yeung_2012, Buaria_2026}, equation~\eqref{eq:slope} predicts that the local slope should steepen from $-3/2$ towards $-2$ as $\Rey_\lambda$ increases, a trend whose verification would require a very broad range of Reynolds numbers (preliminary studies at lower resolutions showed no or a very weak dependence). Finally, the relation is purely statistical: it follows from the definitions and from the one-point statistics of $u$ and $\varepsilon$ alone, and involves no dynamical hypothesis. It is thus of a different nature from the dynamical law of figure~\ref{fig:global}, which requires $\langle u^2 \rangle(t)$ and $\langle \varepsilon \rangle(t)$ to co-evolve locked by the cascade, and measures the scale-by-scale energy imbalance of the flow as a whole rather than one-point statistics.

The local relation can be also read in two complementary ways: as Lagrangian in time, following one fluid element along its trajectory (figure~\ref{fig:local}, left), or as local in space, comparing different fluid elements at a fixed time (figure~\ref{fig:local}, right). Both readings give the same power law with the same exponent: fits at fixed time yield $\alpha = -1.62 \pm 0.01$ (mean and standard deviation over the sampled times), compatible with the per-trajectory value quoted above, and with no discernible dependence on the time considered. This is expected from the interpretation given above: as the relation is fixed by equation~\eqref{eq:identity} and by the one-point statistics of the flow, and as fluid elements sample these statistics ergodically, the cloud of points must be the same whether it is traversed in time or in space. We further verified that the same relation, with the same exponent, holds when the quantities are computed at a fixed time over all grid points of the Eulerian velocity field instead of at the particle positions (not shown), discarding any sampling bias of the tracers. In other words, while the global quantities obey the non-equilibrium law $C_\varepsilon^G \sim (\Rey_\lambda^G)^{-1}$, this law describes the mean state of the flow and only emerges after averaging over the domain. Pointwise, the joint fluctuations of velocity and dissipation are governed by the much broader one-point statistics of the turbulent field, with the steeper local slope. The two exponents thus answer two different questions: \emph{where} a fluid element sits in the intermittent field, and \emph{when} the flow sits in its global cycle. It is then natural to ask how one answer turns into the other as the measurement is progressively averaged, and at which scale the dynamics of the cascade becomes visible.

\section{The emergence of cascade dynamics}

To answer this question we construct a family of quantities that interpolates between local and global definitions. We partition the $N$ tracers into $M = N/K$ randomly chosen, fixed groups of $K$ tracers, and compute for each group
\begin{equation}
	\Rey_\lambda^{(K)} = \sqrt{\frac{15}{\nu \ev{\varepsilon}_K}} \, \ev{u^2}_K \qc
	C_\varepsilon^{(K)} = \frac{\ev{\varepsilon}_K L(t)}{\ev{u^2}_K^{3/2}},
	\label{eq:coarse_defs}
\end{equation}
where $\langle \cdot \rangle_K$ denotes the average over the positions of the $K$ tracers in one group, so that, e.g., $\langle u^2 \rangle_K^{1/2}$ is the r.m.s. velocity sampled by that group. Each group acts as a sensor that reads the flow at $K$ points at once: for $K = 1$, equation~\eqref{eq:coarse_defs} reduces identically to the local quantities of equation~\eqref{eq:local_defs} (as $\langle u^2 \rangle_1 = u^2$), while for $K = N$ it recovers the global quantities of equation~\eqref{eq:global_defs}: $\Rey_\lambda^G = \Rey_\lambda^{(N)}$ and $C_\varepsilon^G = C_\varepsilon^{(N)}$.

\begin{figure}
	\centering
	\includegraphics[width=0.95\linewidth]{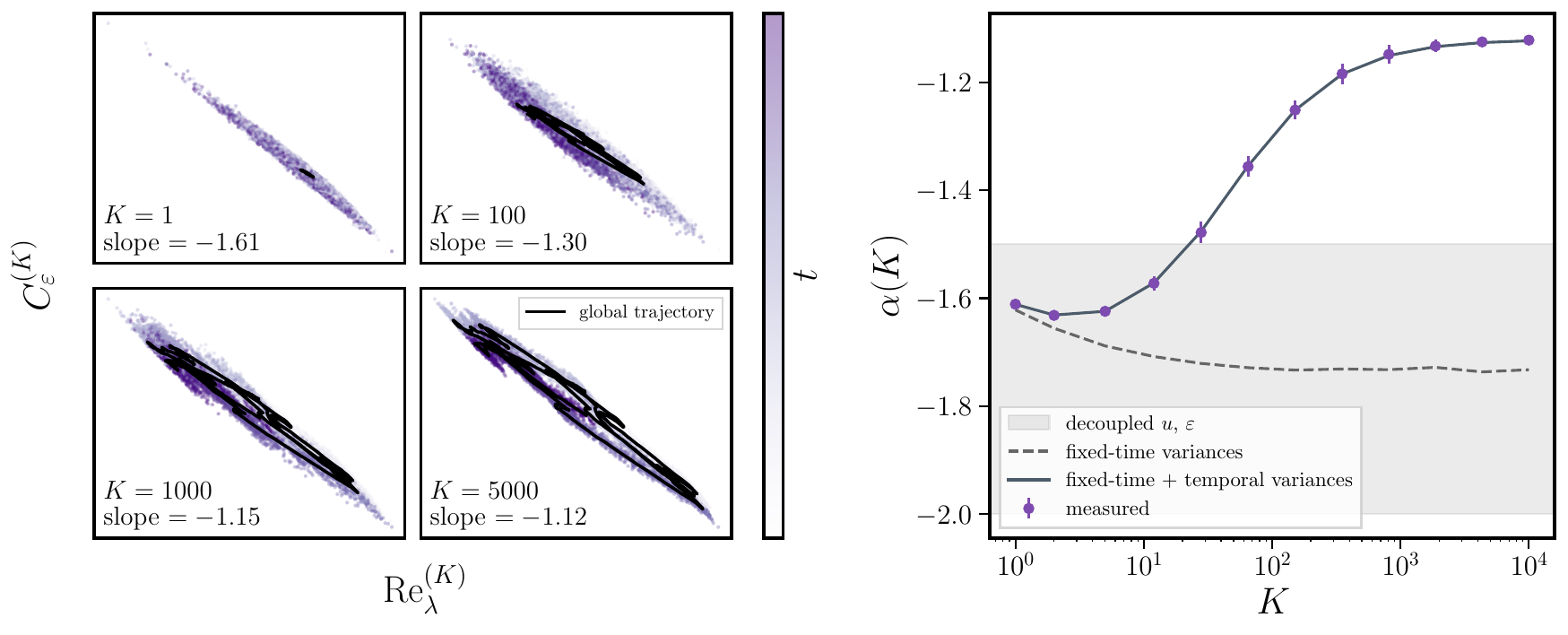}
	\caption{Coarse-grained dimensionless dissipation rate $C_\varepsilon^{(K)}$ as a function of the coarse-grained Reynolds number $\Rey_\lambda^{(K)}$, equation~\eqref{eq:coarse_defs}, and slope $\alpha(K)$ of the relation. The four left panels show the $(\Rey_\lambda^{(K)}, C_\varepsilon^{(K)})$ plane for increasing group sizes $K$ (see labels), with the slope fitted on each cloud reported in the corresponding panel. Each point corresponds to one group at one time, and is coloured by time. The solid line shows the global trajectory of figure~\ref{fig:global} (right), barely visible at the centre of the $K = 1$ cloud; this panel contains the same data as the right panel of figure~\ref{fig:local}, now pooled over all times. The right panel shows $\alpha(K)$ as a function of the group size $K$: circles correspond to the measured slopes, with error bars estimated from a bootstrap over snapshots; the dashed line shows equation~\eqref{eq:slope_general} evaluated with the fixed-time variances and covariances of the group averages, and the solid line shows the two-component prediction of equation~\eqref{eq:alpha_K}, which combines the fixed-time and the temporal variances and covariances. The shaded band marks the interval $[-2, -3/2]$ accessible to decoupled velocity and dissipation fluctuations.}
	\label{fig:transition_alpha}
\end{figure}

Figure~\ref{fig:transition_alpha} (left) shows the $(\Rey_\lambda^{(K)}, C_\varepsilon^{(K)})$ plane for increasing group sizes, where each point corresponds to one group at one time and is coloured by time. The two extreme panels make the two regimes visible. The $K = 1$ panel corresponds to the right panel of figure~\ref{fig:local}, now pooled over all times and coloured by time. Its colours are completely mixed: the fluctuations of a single fluid element carry no information about the phase of the global cycle of figure~\ref{fig:global} (which is shown by the solid black line; note the relatively much larger local fluctuations). Local fluctuations are dominated by the position of the element in the intermittent field, and the $K=1$ cloud follows the local scaling discussed above, with the global trajectory buried near its centre. In the opposite case, for the largest $K$ shown, the sampling scatter has been averaged out and the cloud collapses onto the global trajectory itself, with the time ordering now evident in the colour gradient, including the hysteresis loops associated with the dissipation lag. Note also how the slope fitted on each cloud, reported in the panels, drifts accordingly from the steep local exponent towards the shallower global value of $-1.12$. Increasing $K$ thus acts as a filter: it removes the fluctuations of the intermittent field, which are spatially less coherent and average out between different tracers, while leaving untouched the temporal excursions of the flow as a whole, which are coherent across all tracers.

The transition can be quantified by the slope $\alpha(K)$. Figure~\ref{fig:transition_alpha} (right) shows the measured $\alpha(K)$, together with the prediction obtained by evaluating equation~\eqref{eq:slope_general} with the fixed-time variances and covariances (dashed line). At fixed time, groups differ only by sampling fluctuations; both $\sigma^2_{\ln \langle u^2 \rangle_K}$ and $\sigma^2_{\ln \langle \varepsilon \rangle_K}$ decay as $1/K$, so this prediction depends only on their ratio and stays confined to the range $[-2, -3/2]$. The measured slope starts within this range at $K = 1$ (and at the value theoretically expected), but departs from this prediction (dashed line) after the first few increasing group sizes, leaves the range, and converges to the $\approx -1$ slope of the global law: there the dominant fluctuations are no longer those of the sampling but those of the global cycle, and the slope corresponds to the cascade-locked co-evolution (and cross-correlation) of energy and dissipation, or of $u$ and $\varepsilon$. Equation~\eqref{eq:slope_general} in fact quantifies this, as it is an identity for the slope of any cloud of points in the $(\Rey_\lambda, C_\varepsilon)$ plane, be the cloud built from individual fluid elements, from groups of tracers, or from the flow as a whole, with $\ln u$ and $\ln \varepsilon$ standing for the logarithms of the velocity and dissipation of each cloud. Evaluated with the temporal variances and covariances of the global time series of the flow, it reproduces the global slope with the balance of terms reversed with respect to the local case: the variances alone would give a slope close to $-3/2$, and it is the covariance $C_{u\varepsilon}$ between energy and dissipation, negligible for individual fluid elements but relevant for the flow as a whole, that brings it to $-1$: as the global law amounts to a frozen scale separation $L/\lambda^G$, for constant $L$ one has $\langle \varepsilon \rangle_N \propto \langle u^2 \rangle_N$ along the excursions, i.e., $\sigma^2_{\ln\varepsilon} = 4 \sigma^2_{\ln u}$ and $C_{u\varepsilon} = 2 \sigma^2_{\ln u}$, for which equation~\eqref{eq:slope_general} gives exactly $-1$ (note that covariances involving $L(t)$ still vanish).

Therefore, the behaviour of $\alpha(K)$ has two contributions. A local (spatial at fixed times) contribution which we denote $\alpha_s$ and that is governed by equation~\eqref{eq:slope_general} with the fixed-time variances and covariances (dashed line in figure~\ref{fig:transition_alpha}, right), which, as these covariances are small, is close to equation~\eqref{eq:slope}, and a global contribution which we denote $\alpha_G$ and that is given by the more general equation~\eqref{eq:slope_general} with non-negligible time correlations between $\ln u$ and $\ln \varepsilon$. The weight of each contribution depends on $K$. More specifically, the global contribution is independent of $K$, while the local contribution decreases as $1/K$. The general expression can be obtained from equation~\eqref{eq:slope_general}, but as the intermittent local scatter and the global cycle of the dissipation are uncorrelated, we can write the resulting slope $\alpha(K)$ simply as a variance-weighted average,
\begin{equation}
	\alpha(K) = \frac{\alpha_{\rm s}\, \sigma^2_{\rm s} + \alpha_G\, \sigma^2_{\rm t}}{\sigma^2_{\rm s} + \sigma^2_{\rm t}},
	\label{eq:alpha_K}
\end{equation}
where $\sigma^2_{\rm s} \propto 1/K$ as already mentioned, and $\sigma^2_{\rm t} = \text{Var}_t(\ln \Rey_\lambda^{G})$ is the temporal contribution to the variance of the abscissa $\ln \Rey_\lambda^{(K)}$; equation~\eqref{eq:alpha_K} is in fact equation~\eqref{eq:slope_general}, with each variance and covariance split into its spatial (fixed-time) and temporal parts. Evaluated with the measured variances, equation~\eqref{eq:alpha_K} reproduces the measured slope over the entire range of $K$ (solid line in figure~\ref{fig:transition_alpha}, right). As with equation~\eqref{eq:slope_general}, note that this is a decomposition with no adjustable parameters: although its inputs are measured, they are obtained from the fixed-time statistics and from the global time series separately, neither of which involves the pooled clouds being described. The agreement over the whole range of $K$ supports the assumption that the spatial sampling scatter is statistically independent of the phase of the global cycle. The two variances balance at $\sigma^2_{\rm s}(K^*) = \sigma^2_{\rm t}$, which sets the midpoint of the transition, and gives $K^* \approx 84$ for our time window; as $\sigma^2_{\rm s} \propto 1/K$, the transition is logarithmically broad, spanning about two decades in $K$. Note that, as the groups are random subsets of the tracers spread over the whole domain, $K$ controls the statistical weight of the average rather than a spatial coarse-graining scale, and the group averages converge faster than those of spatially compact sensors: the present $K^*$ is a lower bound for the latter. These results have a direct implication for measurements: depending on the conditions, the non-equilibrium law may not be accessible to sparse instantaneous probes. An observer measuring the flow with a few sensors may not obtain a noisy estimate of the global law, but a clean power law of a different, statistical origin; the dynamical law only emerges once the measurement aggregates enough sensors for the sampling variance to fall below the temporal variance of the global excursions. The variance balance above provides a quantitative criterion. However, the non-equilibrium scaling $C_\varepsilon \sim \Rey_\lambda^{-1}$ has recently been reported to hold at local scales, over cubic subdomains of side $\sim L$ \citep{Nishimoto_2026a, Nishimoto_2026b} or short time windows along single probes \citep{Zheng_2023}, seemingly against the steep local relation found here. Such measurements, however, still average over an entire integral-scale eddy: a region of side $\sim L$ encloses $(L/\eta)^3 \gg K^*$ degrees of freedom and therefore already lies in the coarse-grained regime, so the law they recover is the dynamical one and the statistical relation of the individual fluid element surfaces only at pointwise scales. Similar conditions are met when an external forcing imposes large and slow large-scale excursions as in \citet{Ahmad_2026}, such that global excursions become dominant.

\section{Conclusion}

We have shown that the relation between the dimensionless dissipation rate and the Taylor-scale Reynolds number in a turbulent flow whose global state evolves in time depends on the scale of the sensors that measure it, and that this dependence separates cleanly into two regimes. Considered at the level of individual fluid elements (or, equivalently, of individual points of the Eulerian field), the $(\Rey_\lambda, C_\varepsilon)$ plane displays a steep non-equilibrium local power law with exponent $\approx -1.6$ that contains no global cascade dynamics: it is fixed by the pointwise identity of equation~\eqref{eq:identity} and by the one-point statistics of velocity and dissipation, with the departure from $-3/2$ scaling directly measuring the width of the logarithmic dissipation fluctuations relative to those of the velocity. Considered at the level of the flow as a whole, the same plane displays the non-equilibrium global dissipation law $C_\varepsilon \sim \Rey_\lambda^{-1}$, the signature of the delayed, cascade-locked co-evolution of energy and dissipation. Coarse-graining over groups of $K$ tracers interpolates continuously between the two readings: the measured slope $\alpha(K)$ follows a statistical null model up to a crossover $K^*$, predicted without adjustable parameters by a variance balance, beyond which it leaves the range $[-2, -3/2]$ bounded by decoupled statistics of energy and dissipation, and converges to the global law. This transition is driven by the correlation between velocity and dissipation, which emerges as the number of sampled points grows. For individual fluid elements this correlation is negligible, and velocity and dissipation fluctuate in an effectively decoupled way. Once the sampling fluctuations are averaged out, it becomes the dominant contribution to the slope. This correlation is the signature of the energy cascade, which couples the energy at the large scales with the dissipation at the small scales, and is therefore non-local in scale.

Steep power laws in scatter plots of $C_\varepsilon$ against $\Rey_\lambda$ built from local or instantaneous quantities thus provide information on the intensity of the pointwise fluctuations of dissipation, but not on its dynamics. A power law with exponents between $-3/2$ and $-2$ is the default outcome of uncorrelated local fluctuations between energy and dissipation (but still carries information of how broad the fluctuations are in logarithmic space), with only departures from the statistical prediction carrying dynamical information.

\begin{bmhead}[Funding.]
P.D.M. and B.L.E. acknowledge financial support from UBACyT Grant No. 20020220300122BA, and from Proyecto REMATE of the Redes Federales de Alto Impacto, Argentina.
\end{bmhead}

\begin{bmhead}[Declaration of interests.]
The authors report no conflict of interest.
\end{bmhead}

\begin{bmhead}[Data availability statement.]
The code GHOST used to perform the simulations is publicly available at \url{https://doi.org/10.5281/zenodo.15472741}.
\end{bmhead}

\begin{bmhead}[Author ORCIDs.]
B.L. Espa\~nol, \url{https://orcid.org/0000-0001-7957-7344}; J.C. Vassilicos, \url{https://orcid.org/0000-0003-1828-6628}; R. Zamansky, \url{https://orcid.org/0000-0001-6704-5316}; M. Obligado, \url{https://orcid.org/0000-0003-3834-3941}; P.D. Mininni, \url{https://orcid.org/0000-0001-6858-6755}.
\end{bmhead}

\bibliographystyle{jfm}
\bibliography{bib}

\end{document}